\documentclass[pre,showpacs,amsmath,amssymb]{revtex4}

\usepackage{wasysym}
\usepackage{graphicx}
\usepackage{dcolumn}
\usepackage{bm}
\usepackage{xcolor}
\usepackage{amssymb, amsmath}

\begin{document}

\title{Fokker-Planck equation for transport  of wave packets in nonlinear disordered media}

\author{Nicolas Cherroret and Thomas Wellens}
\affiliation{
Physikalisches Institut, Albert-Ludwigs-Universit\"{a}t Freiburg, Hermann-Herder-Str. 3, D-79104 Freiburg, Germany
}

\date{\today}

\begin{abstract}
Starting from first principles, we formulate a theory of wave packet propagation in a nonlinear, disordered medium of any dimension, through the derivation of a Fokker-Planck transport equation. Our theory is based on a diagrammatic expansion of the wave packet's density, and is supported by a heuristic picture that involves a Boltzmann equation with an effective, external potential. Our approach also confirms results obtained in previous work for two-dimensional, nonlinear disordered media.
\end{abstract}


\maketitle
 
 \section{Introduction}

When it propagates in a disordered environment, a particle (or more generally a wave) usually experiences a classical random walk, with ballistic motion interspersed with a large number of scattering processes \cite{Sheng06}. The overall motion is then described by a diffusion equation for the ensemble-averaged particle's density. In some cases however, this simple picture breaks down due to constructive interference that build up between multiple scattering paths, leading to alteration of the diffusive description. Weak localization, and in the extreme case Anderson localization, which corresponds to a complete halt of transport, are examples of these interference phenomena \cite{Sheng06, Tiggelen99}. Another effect that may strongly affect transport is the possible presence of nonlinearities. In the optical context for instance, the nonlinearity can be obtained by increasing the intensity of the light propagating in the disordered medium, making use of the sensitivity of the refractive index to the electric field (the Kerr effect). Recent works exploited this phenomenon to study the interplay between disorder and nonlinearity in systems of optical fibers \cite{Pertsch04, Lahini08}.

Another promising system for studying quantum transport in disordered environments and in the presence of nonlinearity are ultracold atomic gases. In these systems, the disordered medium is produced by means of an optical random potential, and the nonlinearity originates from interactions between atoms. In this context, the particular problem of a Bose-Einstein condensate prepared in an optical trap and then released in a random potential has sparked considerable interest in the last few years \cite{Fallani08}. The major advantage of this type of experiment is that the time and spatial evolution of the macroscopic wave function of the condensate can be observed by imaging the atomic density profile. Furthermore, experiments involving condensates offer unprecedented control of the physical parameters driving the disorder and the nonlinearity.

Until now, much experimental effort has been devoted to the study of Bose-Einstein condensates in one-dimensional random potentials, with strong interest in the phenomenon of Anderson localization, the role of atomic interactions, and the interplay between these effects \cite{Modugno10}. In recent works, Anderson localization of a condensate was experimentally demonstrated, in a situation where interactions were presumably negligible \cite{Billy08, Roati08}. Although one might expect that interactions play no role after a long period of expansion of the condensate, during which it becomes increasingly dilute, recent theoretical works suggested that, even at long times, Anderson localization could be destroyed by interactions, leading to a subdiffusive regime \cite{Pikovsky08, Flach09, Lucioni11}.
 
In higher dimensions, our knowledge of the behavior of Bose-Einstein condensates in the presence of disorder is still in its fledgling stage. Among the handful of experiments that have been carried out, one can cite recent beautiful works involving ultracold atoms, in which the Anderson transition was demonstrated and analyzed in detail, through the experimental realization of a quantum-chaotic system, the quasiperiodic kicked rotor, known to be equivalent to the three-dimensional disordered system \cite{Chabe08, Lemarie10}. Regarding the quantum transport of Bose-Einstein condensates in random potentials, the common feature between the two- and three-dimensional cases is that unlike in one dimension (1D), a broad diffusion regime exists, provided the disorder is not too strong. Here, ``diffusion regime" means that most atoms of the condensate propagate by diffusion. In three dimensions (3D), it was shown that this situation is achieved if the quantity $\mu\tau/\hbar$ is large \cite{Cherroret09}, where $\mu$ is the chemical potential of the condensate and $\tau$ the scattering time, \emph{i.e.}, the average time between two successive scattering processes. In 3D, both the diffusion \cite{Shapiro07} and the localization \cite{Skipetrov08} regimes have been studied theoretically, as well as the cross-over between them \cite{Cherroret09}. From the experimental point of view, the diffusion of a cloud of thermal atoms in a random potential was recently studied \cite{Robert10}. Yet, in two and three dimensions, whether it concerns a regime where localization effects dominate or a situation where transport is driven by diffusion, the role of atomic interactions on condensate expansion remains largely unknown. Furthermore, to our knowledge no analytical description of this type of problem is available. In a recent theoretical work however, a kinetic equation describing the propagation of a wave packet in the infinite two-dimensional disordered medium was proposed, based on the introduction of an effective nonlinear potential \cite{Schwiete10}. From this equation, the authors of Ref. \cite{Schwiete10} were able to infer the qualitative behavior of the wave packet in the presence of nonlinearity, and, in particular, put forward the phenomenon of ``locked" explosion, corresponding to a global diffusive behavior, but with a rapid explosion at early stages of the expansion, located in the central part of the wave packet.

This paper aims at formulating a consistent theory able to describe wave packet propagation in a nonlinear, weakly disordered medium. For this purpose, we derive a Fokker-Planck equation describing transport of such a wave packet, starting from first principles. 
This equation holds in any dimension, and, in 2D, reduces to the kinetic equation introduced in Ref. \cite{Schwiete10}. Our approach is based on a diagrammatic treatment of the wave packet's density in the presence of nonlinearity. Such a formalism was initially introduced to describe the stationary transport of waves in nonlinear media either formed by nonlinear point scatterers at random positions \cite{Wellens08, Wellens09} or by linear scatterers embedded in a homogeneous nonlinear medium \cite{WellensApll}. In particular, it was successfully applied to the description of the coherent backscattering effect in the presence of nonlinearities \cite{Hartung08}. In this paper, we extend this approach to time dependent diffusion processes, and, in addition, validate the diagrammatic approach by a heuristic treatment that consists in solving the Boltzmann equation in the presence of an effective, nonlinear potential, in the spirit of \cite{Schwiete10}. We work in the framework of quantum particles, but our results can be extended to classical wave scattering described by the nonlinear Helmholtz equation with a disordered potential \cite{WellensApll}.

The paper is organized as follows. In Sec. II, we introduce the physical quantities relevant to describe transport of a wave packet in a nonlinear disordered medium. We introduce the Fokker-Planck equation and establish the connection with the work in Ref. \cite{Schwiete10}. The two main approximations used throughout the paper, the diffusion approximation and the limit of weak disorder, are discussed, and the validity of the latter in a nonlinear medium is analyzed in Sec. III. In Sec. IV, we recall the core elements of the derivation of the standard diffusion equation describing wave packet dynamics in a linear disordered medium. This derivation is extended to the nonlinear case in Secs. \ref{section5} and \ref{section6}, leading to the Fokker-Planck equation. Finally, in Sec. \ref{section7}, we show how this equation can be straightforwardly obtained from a heuristic approach resorting to the Boltzmann equation in the presence of an external, effective potential describing the nonlinearity. Technical details of calculations are collected in the three appendices.

 \section{Theoretical framework}
 \label{section2}
 
Before examining the nonlinear case, we first recall some basic facts concerning the propagation of a wave packet in the infinite, linear disordered medium of dimension $d$ ($d=1$, $2$ or $3$). This wave packet can represent, for instance, a collection of particles of mass $m$ moving in a disordered environment. At a given point $\textbf{r}$ inside the medium and at time $t$, the wave function of this wave packet obeys the time-dependent Schr\"{o}dinger equation:
 \begin{equation}
 \label{Schro}
 i\dfrac{\partial\psi(\textbf{r},t)}{\partial t}=\left[-\dfrac{1}{2m}\boldsymbol{\nabla}^2+V(\textbf{r})\right]\psi(\textbf{r},t).
 \end{equation}
In Eq. (\ref{Schro}), and in the rest of the paper, we set $\hbar=1$. The disordered medium is described by the random potential $V(\textbf{r})$, that we assume to obey the white-noise Gaussian statistics
\begin{equation}
\label{correlation}
\overline{V(\textbf{r})V(\textbf{r}^\prime)}=\gamma\delta(\textbf{r}-\textbf{r}^\prime),
\end{equation}
where the overbar denotes averaging over realizations of disorder and $\gamma=1/(2\pi\nu\tau)$, with $\tau$ the scattering time and $\nu=(d/2)(m/2\pi)^{d/2}\epsilon^{d/2-1}/\Gamma(1+d/2)$ the density of states per unit volume. By Fourier transforming Eq. (\ref{Schro}) with respect to time we obtain
\begin{equation} \label{Schro_Fourier}
\left[-\epsilon-\dfrac{1}{2m}\boldsymbol{\nabla}^2+V(\textbf{r})\right]\psi_{\epsilon}(\textbf{r})=0,
\end{equation}
where $\psi_{\epsilon}(\textbf{r})=\int dt\, \psi(\textbf{r},t)e^{i\epsilon t}$. In this paper, we will be interested in the average density at energy $\epsilon$, defined as
\begin{equation}
n_\epsilon(\textbf{r},t)= \int \dfrac{d\omega}{2\pi}\, n_\epsilon(\textbf{r},\omega)e^{-i\omega t},
\end{equation}
where $n_\epsilon(\textbf{r},\omega)=\overline{\psi_{\epsilon+\omega/2}(\textbf{r})\psi_{\epsilon-\omega/2}^*(\textbf{r})}$. Physically, $n_\epsilon(\textbf{r},t)$ can be interpreted as the probability density of finding a particle with energy $\epsilon$ in the vicinity of point $\textbf{r}$, at time $t$. Even in the absence of nonlinear effects, the evaluation of this object is, in general, a difficult task that requires to average over the product of the sum of all partial waves reaching the observation point $\textbf{r}$. The analysis is, however, greatly simplified if one restricts to the so-called diffusion approximation and to the limit of weak disorder \cite{AM}. In the diffusion approximation, one considers only large spatial scales $|\textbf{r}|\gg\ell$ (assuming the wave packet is initially located at point \textbf{r = 0}), where $\ell$ is the mean free path, and slow dynamics $\omega\ll\epsilon,1/\tau$. The limit of weak disorder corresponds to the condition $\epsilon\tau\gg1$ and describes complete phase  randomization of the partial fields, such that no interference survives the disorder average \cite{AM}.  ``Phase coherent" effects such as weak or strong (Anderson) localization are thus neglected in this picture. Under these assumptions, the average density obeys the diffusion equation \cite{AM}
\begin{equation}
\label{diff_eq}
(\partial_t-D_\epsilon\boldsymbol{\nabla}^2)n_\epsilon(\textbf{r},t)=\delta(t)n_\epsilon(\textbf{r},t=0).
\end{equation}
The right-hand side of this equation is the ``source term", described as a short pulse whose magnitude is given by the wave packet's density at the initial time. The main lines of the derivation of Eq. (\ref{diff_eq}) will be recalled in Sec. \ref{section4}. $D_\epsilon=v^2\tau/d$ is the diffusion coefficient at energy $\epsilon$ in dimension $d$, expressed as a function of the scattering time and the velocity $v=\sqrt{2\epsilon/m}$. From the solution $n_\epsilon(\textbf{r},t)$ of Eq. (\ref{diff_eq}), the \emph{total} density of the wave packet is obtained after integration over energies:
\begin{equation}
n(\textbf{r},t)\equiv |\psi(\textbf{r},t)|^2=\int \dfrac{d\epsilon}{2\pi}n_\epsilon(\textbf{r},t).
\end{equation}

Let us now introduce nonlinearity in the medium. If this nonlinearity is ``not too strong", the wave function of the wave packet obeys the Gross-Pitaevskii equation
 \begin{equation}
 \label{GPE}
 i\dfrac{\partial\psi(\textbf{r},t)}{\partial t}=\left[-\dfrac{1}{2m}\boldsymbol{\nabla}^2+V(\textbf{r})+g|\psi(\textbf{r},t)|^2 \right]\psi(\textbf{r},t).
 \end{equation}
The term $g|\psi|^2$ plays the role of a nonlinear potential, repulsive if the parameter $g$ is positive, and attractive if $g$ is negative. In the context of ultracold atomic gases, the Gross-Pitaevskii equation controls the time-evolution of a Bose-Einstein condensate, described in terms of a macroscopic wave function $\psi$, in the presence of a random potential $V(\textbf{r})$. The nonlinearity then originates from interactions between atoms \cite{Pitaevskii03}. 

The presence of the nonlinear term in Eq. (\ref{GPE}) substantially complicates the analysis of the problem. In this paper, we show that in a nonlinear, disordered medium, the diffusion equation (\ref{diff_eq}) for the average density $n_\epsilon(\textbf{r},t)$ must be replaced by a Fokker-Planck equation that reads
\begin{equation}
\label{FP_eq}
\left[\partial_t -\boldsymbol{\nabla}D_{\epsilon-\theta}\boldsymbol{\nabla}+\partial_t \theta\partial_\epsilon +\boldsymbol{\nabla}\eta_{\epsilon-\theta}\textbf{F}\right]n_\epsilon(\textbf{r},t)=\delta(t)n_{\epsilon-\theta}(\textbf{r},t=0),
\end{equation}
where we have introduced the ``nonlinear potential" $\theta=\theta(\textbf{r},t)=2gn(\textbf{r},t)$ and its associated ``force" $\textbf{F}=-\boldsymbol{\nabla}\theta$. Eq. (\ref{FP_eq}) is valid in any dimension, and thereby generalizes the kinetic equation analyzed in \cite{Schwiete10}, which only applies to the two-dimensional case (for the sake of clarity we have kept the same notations as in \cite{Schwiete10}). 

Although Eq. (\ref{FP_eq}) formally describes the propagation of a wave packet in a disordered, nonlinear environment, its analysis reveals that the problem is analogous to the one of a collection of independent particles, moving in a disordered region and subject to the external force $\textbf{F}=-\boldsymbol{\nabla}\theta$. Each particle has total energy $\epsilon$ and kinetic energy $\epsilon-\theta$. The parameter $\eta_{\epsilon-\theta}=(d/2-1)\mu_{\epsilon-\theta}$ depends on the dimensionality $d$ and on the quantity $\mu_{\epsilon-\theta}=D_{\epsilon-\theta}/(\epsilon-\theta)$, which can be seen as the mobility of a particle with kinetic energy $\epsilon-\theta$. The main consequence of the presence of the potential $\theta$ is that the diffusion coefficient, which depends on the kinetic energy $\epsilon-\theta$, now acquires a position and time dependence.  The third term in the left-hand side of Eq. (\ref{FP_eq}) originates from the time dependence of the potential $\theta$, and the last term corresponds to a drift effect. For $d=2$, we have $\eta_\epsilon=0$ and one recovers the kinetic equation analyzed in \cite{Schwiete10}. Although Eq. (\ref{FP_eq}) is rather complicated, we can gain more physical insight by performing the substitution $\epsilon_0=\epsilon-\theta$, and integrating the resulting equation over $\epsilon_0$. We then obtain the following simple continuity equation for the total density:
\begin{equation}
\label{continuity}
\partial_t n(\textbf{r},t)+\boldsymbol{\nabla}\cdot\textbf{J}(\textbf{r},t)=\delta(t)n(\textbf{r},t=0),
\end{equation}
where the total current $\textbf{J}$ is defined as
\begin{equation}
\label{total_J}
\textbf{J}(\textbf{r},t)=\int\dfrac{d\epsilon_0}{2\pi}\left[-D_{\epsilon_0}\boldsymbol{\nabla}n_{\epsilon_0}(\textbf{r},t)+\mu_{\epsilon_0}\textbf{F}n_{\epsilon_0}(\textbf{r},t)\right].
\end{equation}
The continuity equation guarantees the conservation of the total density $n(\textbf{r},t)$. The formula (\ref{total_J}) has a clear physical interpretation: the total current is given by the sum of a diffusion current (present in the linear case), and a drift current, induced by the force $\textbf{F}$. This drift current is proportional to the mobility $\mu_{\epsilon_0}=D_{\epsilon_0}/\epsilon_0$. Again, Eq. (\ref{total_J}) holds in any dimension. In the particular two-dimensional case, $D_{\epsilon}\propto\epsilon$, such that Eqs. (\ref{continuity}) and (\ref{total_J}) can be simplified as
\begin{equation}
\label{total_J_2d}
\partial_t n(\textbf{r},t)=\dfrac{\tau}{m}\boldsymbol{\nabla}^2\left[\int\dfrac{d\epsilon_0}{2\pi}\epsilon_0n_{\epsilon_0}(\textbf{r},t)+gn^2(\textbf{r},t)\right],
\end{equation}
for $t>0$. From the particular simple form of Eq. (\ref{total_J_2d}), it follows that the mean radius of the wave packet grows as $t^{1/2}$ in 2D, similarly as in the linear case. This phenomenon, named ``locked explosion" in Ref. \cite{Schwiete10}, is a consequence of the fact that the mobility $\mu_{\epsilon_0}$ is independent of $\epsilon_0$ in 2D. In 1D or 3D, we expect different physical scenarios because the mobility acquires an energy dependence. This is however still an open question that we leave for later work.

The Fokker-Planck equation (\ref{FP_eq}) is a promising tool to study transport in nonlinear disordered media, and therefore calls for microscopic justification. Derivation of this equation from first principles and the formulation of a consistent theory of dynamical transport in a nonlinear, disordered medium, are the two purposes of the present contribution. In the following, we will focus on the three-dimensional infinite medium, and explain how results are modified in 1D and 2D.

 \section{Limit of weak disorder in the presence of nonlinearity}
 \label{section3}

In the microscopic approach developed hereafter, we will make use of the limit of weak disorder $\epsilon\tau\gg1$ discussed in Sec. \ref{section2}, assuming that this condition is not modified in the presence of nonlinearity. This assumption amounts to neglect the effect of the nonlinearity at the scale of the scattering time, and leads to a restriction for the values taken by the average density. To explicit this restriction, we express that the scattering time $\tau_{\text{NL}}$ associated with the nonlinear potential $gn(\textbf{r},t)$ in Eq. (\ref{GPE}) should be much larger than the scattering time $\tau$ associated with the disordered potential \cite{Spivak00, Skipetrov00}. To calculate $\tau_{\text{NL}}$, we use the short-range ``$C_1$" part of the spatial correlation $\overline{ \delta n_\epsilon(\textbf{r},t)\delta n_\epsilon(\textbf{r}^\prime,t)}$ of density fluctuations \cite{AM}. The condition $\tau_\text{NL}\gg\tau$ then leads to:
\begin{equation}
\label{cond_tau}
\tau g n(\textbf{r},t)\ll\sqrt{\epsilon\tau}.
\end{equation}
Condition (\ref{cond_tau}) will be assumed to hold from here on. Note that to obtain it, we neglected the long-range ``$C_2$"part of the correlation function of density fluctuations \cite{AM}. This approximation is justified in our case, since one readily shows that at long times, $C_2\propto (1/t^{d-1})/(k\ell)^{d-1}$ in dimension $d$ \cite{Cherroret09}. The time decay of $C_2$ then allows us to neglect its contribution to $\tau_{\text{NL}}$ in 2D and 3D.  In 1D, this assumption is questionable because $C_2\sim C_1$. This is related to the fact that in 1D, localization effects take place at very short time scales $t\sim\tau$, which, strictly speaking, limits the validity of the diffusive approach. We will come back to this point in Secs. \ref{section6} and \ref{section8}.

\section{Diagrammatic approach in the linear case}
\label{section4}

In this section we recall the derivation of the usual diffusion equation (\ref{diff_eq}) in a linear, disordered medium. The microscopic description of transport is based on a Feynmann representation of the density. More precisely, in the limit of weak disorder, this description is a perturbation theory which consists in finding the leading diagrams in $1/(\epsilon\tau)$ contributing to $n_\epsilon(\textbf{r},\omega)=\overline{\psi_{\epsilon_1}(\textbf{r})\psi_{\epsilon_2}^*(\textbf{r})}$, where $\epsilon_1=\epsilon+\omega/2$ and $\epsilon_2=\epsilon-\omega/2$. These diagrams are formed by combining a scattering sequence associated with $\psi_{\epsilon_1}$ and a scattering sequence associated with $\psi^*_{\epsilon_2}$. The key point is that in the limit of weak disorder, the only diagram that survives the disorder average is the so-called ``ladder" diagram, for which $\psi_{\epsilon_1}$ and $\psi^*_{\epsilon_2}$ follow \emph{exactly} the same scattering sequence, thereby introducing no phase difference between $\psi_{\epsilon_1}$ and $\psi^*_{\epsilon_2}$. The ladder diagram is depicted in Fig. \ref{diagrams}a. It connects the scattering sequence associated with $\psi_{\epsilon_1}$ (solid line) with the one associated with $\psi^*_{\epsilon_2}$ (dashed line). Inner dotted lines symbolize the correlation function of disorder (\ref{correlation}) between a scattering event experienced by  $\psi_{\epsilon_1}$ and one experienced by  $\psi^*_{\epsilon_2}$. The ladder diagram is generated by the ``building block" depicted in Fig. \ref{diagrams}b, whose mathematical representation is the Bethe-Salpeter equation \cite{AM}:
 \begin{figure}[h]
\includegraphics[width=12cm]{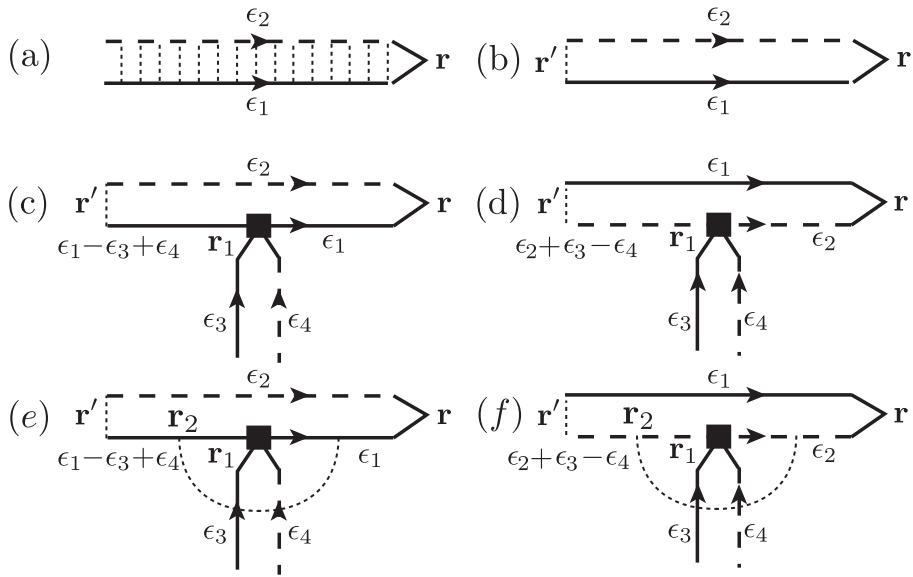}
\caption{\label{diagrams} 
(a) Ladder diagram, connecting the scattering path associated with by $\psi_{\epsilon_1}$ (solid line) to the scattering path associated with $\psi^*_{\epsilon_2}$ (dashed line). $\textbf{r}$ is the observation point. Arrows indicate the direction of propagation. Inner dotted lines symbolize the correlation function of disorder given by Eq. (\ref{correlation}). (b) Building block generating the ladder diagram, and leading to the Bethe-Salpeter equation (\ref{BS_eq}). The solid and dashed lines depict Green's functions $\overline{G}_{\epsilon_1}$  and $\overline{G}^*_{\epsilon_2}$, respectively. The last scattering process of the ladder diagram occurs at point $\textbf{r}^\prime$. (c), (d), (e) and (f) Building blocks generating the contributions $\overline{\psi_{\epsilon_1}(\textbf{r})\psi^*_{\epsilon_2}(\textbf{r})}|_\text{c}$, $\overline{\psi_{\epsilon_1}(\textbf{r})\psi^*_{\epsilon_2}(\textbf{r})}|_\text{d}$,  $\overline{\psi_{\epsilon_1}(\textbf{r})\psi^*_{\epsilon_2}(\textbf{r})}|_\text{e}$ and  $\overline{\psi_{\epsilon_1}(\textbf{r})\psi^*_{\epsilon_2}(\textbf{r})}|_\text{f}$ to the density, respectively. The black square represents the nonlinear parameter $g$. Other symbols have the same meaning as in diagrams (a) and (b).}  
\end{figure}
\begin{equation}
\label{BS_eq}
\overline{\psi_{\epsilon_1}(\textbf{r})\psi^*_{\epsilon_2}(\textbf{r})}=\overline{\psi}_{\epsilon_1}(\textbf{r})\overline{\psi}^*_{\epsilon_2}(\textbf{r})+\gamma\int d^3\textbf{r}^\prime\,\overline{G}_{\epsilon_1}(\textbf{r}^\prime,\textbf{r})\overline{G}^*_{\epsilon_1}(\textbf{r}^\prime,\textbf{r})\overline{\psi_{\epsilon_1}(\textbf{r}^\prime)\psi^*_{\epsilon_2}(\textbf{r}^\prime)},
\end{equation}
where $\overline{G}_\epsilon(\textbf{r}^\prime,\textbf{r})$ is the average ``amplitude Green's function", which is the Green's function of the Schr\"{o}dinger equation (\ref{Schro}), averaged over disorder. In Fig. \ref{diagrams}b, $\overline{G}_{\epsilon_1}$ and $\overline{G}^*_{\epsilon_2}$ are depicted by the solid and dashed lines, respectively. The Bethe-Salpeter is an integral equation for the density. In the diffusion approximation and in the limit of weak disorder, it can be simplified to yield the diffusion equation. For this purpose, it is convenient to work in Fourier space. We thus introduce $n_\epsilon(\textbf{q},\omega)=\int d^3\textbf{r}\,\overline{\psi_{\epsilon_1}(\textbf{r})\psi^*_{\epsilon_2}(\textbf{r})}e^{-i\textbf{q}\cdot\textbf{r}}$ and $S_\epsilon(\textbf{q},\omega)=\int d^3\textbf{r}\,\overline{\psi}_{\epsilon_1}(\textbf{r})\overline{\psi}^*_{\epsilon_2}(\textbf{r})e^{-i\textbf{q}\cdot\textbf{r}}$. A calculation detailed in Appendix \ref{appendix_A} then gives
\begin{equation}
\label{BS_eq_simplified}
n_\epsilon(\textbf{q},\omega)\simeq
S_\epsilon(\textbf{q},\omega)+
(1+i\omega\tau-D_\epsilon\tau\textbf{q}^2)\,
n_\epsilon(\textbf{q},\omega)
\end{equation}
where $D_\epsilon=v^2\tau/d$ is the diffusion coefficient at energy $\epsilon$. Taking the inverse Fourier transform of Eq. (\ref{BS_eq_simplified}) with respect to $\omega$ and $\textbf{q}$, we obtain
\begin{equation}
\label{diff_eq_2}
(\partial_t-D_\epsilon\boldsymbol{\nabla}^2)n_\epsilon(\textbf{r},t)=\dfrac{S_\epsilon(\textbf{r},t)}{\tau},
\end{equation}
which is nothing but a diffusion equation with  source term $S_\epsilon(\textbf{r},t)/\tau$. A calculation detailed in Appendix \ref{appendix_C} leads to
\begin{equation}
\label{source_term_2}
S_\epsilon(\textbf{r},t)=\tau\delta(t)n_\epsilon(\textbf{r},t=0),
\end{equation}
where, in the limit of weak disorder and within the diffusion approximation, the wave packet's density $n_\epsilon(\textbf{r},t=0)$ at the initial time is given by (see Appendix  \ref{appendix_C})
\begin{equation}
\label{source_term}
n_\epsilon(\textbf{r},t=0)\simeq\int\!\!\!\int \dfrac{d^3\textbf{q}}{(2\pi)^3}\dfrac{d^3\textbf{Q}}{(2\pi)^3}\, 2\pi A(\epsilon,\textbf{Q})
\phi(\textbf{Q}+\textbf{q}/2)\phi^*(\textbf{Q}-\textbf{q}/2)e^{-i\textbf{q}\cdot\textbf{r}}.
\end{equation}
In Eq. (\ref{source_term}), $A(\epsilon,\textbf{Q})=-\text{Im}\overline{G}_\epsilon(\textbf{Q})/\pi$ is the ``spectral function", which is the density probability of finding a particle with energy $\epsilon$ among all particles with momentum $\textbf{Q}$ \cite{Skipetrov08}. In the limit of weak disorder $\epsilon\tau\gg1$, the spectral function becomes very close to the free-space expression $A(\epsilon,\textbf{Q})\simeq\delta(\epsilon-\epsilon_Q)$, where $\epsilon_Q=\textbf{Q}^2/(2m)$ \cite{Shapiro07, Schwiete10}. $\phi(\textbf{k})$ is the Fourier transform of the wave function $\phi(\textbf{r})$ of the wave packet at the initial time. The integral over $\textbf{q}$ accounts for the possible finite extent of the wave packet. For the particular case of a ``point source", $\phi(\textbf{r})\propto\delta(\textbf{r})$ and consequently $S_\epsilon(\textbf{r},t)\propto\delta(\textbf{r})\delta(t)$. In this case, $n_\epsilon$ reduces to the Green's function of the diffusion equation, sometimes called ``probability of quantum diffusion" in the literature \cite{AM}.

\section{Diagrammatic approach in the nonlinear case}
\label{section5}

\subsection{Nonlinear diagrams}

\begin{figure}[h]
\includegraphics[width=8cm]{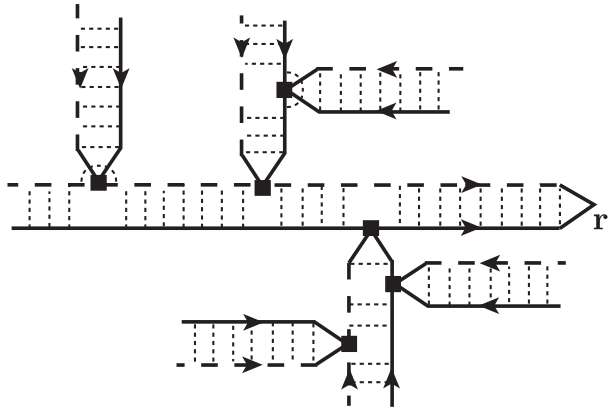}
\caption{\label{general_diagram} 
A typical series of diagrams contributing to the density in the nonlinear case. All ladder diagrams constituting this series originate from a point of the initial wave packet, and are connected to another ladder diagram via two average amplitude Green's functions. This connection is referred to as a nonlinear scattering process in the main text. Only the central diagram is connected to the observation point \textbf{r}.}  
\end{figure}

In the previous section, we saw that in the linear case, only the ladder diagram contributes to the density in the limit of weak disorder. In the nonlinear case, each of the two scattering paths building the ladder diagram may be ``perturbed" by the nonlinear potential $g|\psi|^2$ in Eq. (\ref{GPE}), at any point of the scattering sequence. The density at some point $\textbf{r}$ is then obtained by summing an infinite series of diagrams of the type of the one depicted in Fig. \ref{general_diagram}. In this figure, every ladder diagram originates from a point of the wave packet at the initial time, and is connected to one of the two scattering paths forming another ladder diagram, via two average amplitude Green's functions. From here on, we will refer to such a connection as a ``nonlinear scattering process". In Fig. \ref{general_diagram}, the black square at each connection symbolizes the interaction parameter $g$. More details about this construction can be found in \cite{Wellens09}. 

Since an infinite number of diagrammatic sequences like the one of Fig. \ref{general_diagram} must be accounted for in order to calculate the density, writing down an integral equation similar to the Bethe-Salpeter equation of the linear case seems to be a difficult task. This is however possible if we notice that all possible diagrammatic series are composed of the only five ``building blocks" depicted in Figs. \ref{diagrams}b, \ref{diagrams}c, \ref{diagrams}d, \ref{diagrams}e and \ref{diagrams}f. The density is obtained by summing the integral equations associated with each of these building blocks. This strategy was initially developed in the particular case of a finite nonlinear disordered medium under plane wave excitation, \emph{i.e.}, corresponding to $\omega=0$ (or equivalently $ \epsilon_1=\epsilon_2$) \cite{Wellens09}. In the present case, an additional difficulty proceeds from the energy exchange arising at each nonlinear scattering process. This exchange has to be examined carefully when writing down the integral equations for diagrams \ref{diagrams}b, \ref{diagrams}c, \ref{diagrams}d, \ref{diagrams}e and \ref{diagrams}f. It can be readily understood from the Gross-Pitaevskii equation (\ref{GPE}) which, in Fourier space, can be rewritten as \cite{Cherroret09}
\begin{equation} \label{GPE_Fourier}
\left[-\epsilon-\dfrac{\hbar^2}{2m}\boldsymbol{\nabla}^2+V(\textbf{r})\right]\psi_{\epsilon}(\textbf{r})+g\int\!\!\!\int\dfrac{d\epsilon_1}{2\pi}\dfrac{d\epsilon_2}{2\pi}\psi_{\epsilon_1}(\textbf{r})\psi_{\epsilon_2}^*(\textbf{r})\psi_{\epsilon-\epsilon_1+\epsilon_2}(\textbf{r})=0.
\end{equation}
The nonlinear term in the left-hand side of Eq. (\ref{GPE_Fourier}) has a straightforward interpretation: when propagating in the disordered medium, the partial field $\psi_{\epsilon-\epsilon_1+\epsilon_2}$ is affected by the density $\psi_{\epsilon_1}\psi^*_{\epsilon_2}$, from which an amount of energy $\epsilon_1-\epsilon_2$ is transferred. In general, this energy exchange may affect either the field $\psi$ or its complex conjugate $\psi^*$. This is manifested by the existence of the two building blocks in Figs. \ref{diagrams}d and \ref{diagrams}f, respectively obtained from the building blocks in Figs. \ref{diagrams}c and \ref{diagrams}e, after the substitutions $\overline{G}\leftrightarrow\overline{G}^*$, $\epsilon_1\leftrightarrow\epsilon_2$ and $\epsilon_3\leftrightarrow\epsilon_4$.

Before proceeding further, two important comments are in order. First, keeping \emph{all} the building blocks in Figs. \ref{diagrams}b, \ref{diagrams}c, \ref{diagrams}d, \ref{diagrams}e and \ref{diagrams}f is crucial in order to account properly for nonlinear effects, while preserving conservation of $n(\textbf{r},t)$. Second, although the present approach sums infinitely many diagrams, one has to keep in mind that it remains perturbative since we do not consider situations where more than one nonlinear scattering process occurs in a row. This means that we implicitly assume that the parameter $\tau gn(\textbf{r},t)$ is small, such that condition (\ref{cond_tau}) is automatically fulfilled in the limit of weak disorder $\epsilon\tau\gg1$. As we will discuss in Sec. \ref{section7} however, the Fokker-Planck equation (\ref{FP_eq}) resulting from the diagrammatic approach is valid beyond this assumption.

\subsection{Calculation of building blocks}

We now examine the integral equations corresponding to the building blocks in Figs. \ref{diagrams}c and \ref{diagrams}d. The one in Fig. \ref{diagrams}c generates a series of diagrams for which the scattering sequence associated with $\psi_{\epsilon_1}$ emerges at point $\textbf{r}$ from a \emph{nonlinear} scattering process, and the scattering sequence of $\psi^*_{\epsilon_2}$ from a \emph{linear} scattering process. From here on we denote by $\overline{\psi_{\epsilon_1}(\textbf{r})\psi^*_{\epsilon_2}(\textbf{r})} |_\text{c}$ the contribution to the density associated with this series of diagrams. It obeys  (see Fig. \ref{diagrams}c):
\begin{equation}
\label{BS_eq_c}
\overline{\psi_{\epsilon_1}(\textbf{r})\psi^*_{\epsilon_2}(\textbf{r})} |_\text{c}=
2\gamma g\int\!\!\!\int d^3\textbf{r}^\prime d^3\textbf{r}_1
\int\!\!\!\int \dfrac{d\epsilon_3}{2\pi}\dfrac{d\epsilon_4}{2\pi}
\overline{G}^*_{\epsilon_2}(\textbf{r}^\prime,\textbf{r})
\overline{G}_{\epsilon_1-\epsilon_3+\epsilon_4}(\textbf{r}^\prime,\textbf{r}_1)
\overline{G}_{\epsilon_1}(\textbf{r}_1,\textbf{r})
\overline{\psi_{\epsilon_3}(\textbf{r}_1)\psi^*_{\epsilon_4}(\textbf{r}_1)}\times
\overline{\psi_{\epsilon_1-\epsilon_3+\epsilon_4}(\textbf{r}^\prime)\psi^*_{\epsilon_2}(\textbf{r}^\prime)}.
\end{equation}
Note that the densities that appear inside the integrals, and evaluated at points $\textbf{r}^\prime$ and $\textbf{r}_1$, must correspond to the \emph{total} density in order to effectively sum all diagrams. The prefactor $2$ accounts for two possibilities to group the incoming Green's functions converging at point $\textbf{r}_1$ into pairs (see Fig. \ref{diagrams}c). Similarly, the building block of Fig. \ref{diagrams}d generates a series of diagrams for which the scattering sequence associated with $\psi_{\epsilon_1}$ emerges at point $\textbf{r}$ from a \emph{linear} scattering process, and the scattering sequence associated with $\psi^*_{\epsilon_2}$ from a \emph{nonlinear} scattering process. The associated series of diagrams $\overline{\psi_{\epsilon_1}(\textbf{r})\psi^*_{\epsilon_2}(\textbf{r})} |_\text{d}$ obeys the same equation as $\overline{\psi_{\epsilon_1}(\textbf{r})\psi^*_{\epsilon_2}(\textbf{r})} |_\text{c}$, but with the substitutions $\overline{G}\leftrightarrow\overline{G}^*$, $\epsilon_1\leftrightarrow\epsilon_2$ and $\epsilon_3\leftrightarrow\epsilon_4$.
In the diffusion approximation and in the limit of weak disorder, the integral equations for the diagrams  \ref{diagrams}c and \ref{diagrams}d can be simplified. This calculation is lengthy, and is reported to Appendix \ref{appendix_B} for clarity. It can be conveniently performed in Fourier space, where it leads to
\begin{eqnarray}
\label{BS_eq_cd}
n_\epsilon(\textbf{q},\omega)|_\text{c}&+&
n_\epsilon(\textbf{q},\omega)|_\text{d}=\nonumber\\
&&4g\tau\int\!\!\!\int \dfrac{dE}{2\pi}\dfrac{d\Omega}{2\pi}
\int\dfrac{d^3\textbf{q}_1}{(2\pi)^3}
n_E(\textbf{q}_1,\Omega)
n_\epsilon(\textbf{q}-\textbf{q}_1,\omega-\Omega)
\left[-\dfrac{1+i(\omega-\Omega)\tau}{2k\ell}+\dfrac{\ell^2}{2k\ell}(\textbf{q}^2-\textbf{q}\cdot\textbf{q}_1)\right]+\nonumber\\
&&+2ig\tau\int\!\!\!\int \dfrac{dE}{2\pi}\dfrac{d\Omega}{2\pi}
\int\dfrac{d^3\textbf{q}_1}{(2\pi)^3}
\Omega\, n_E(\textbf{q}_1,\Omega)
\partial_\epsilon n_\epsilon(\textbf{q}-\textbf{q}_1,\omega-\Omega),
\end{eqnarray}
where $k=\sqrt{2m\epsilon}$, $n_\epsilon(\textbf{q},\omega)=\int d^3\textbf{r}\, \overline{\psi_{\epsilon_1}(\textbf{r})\psi^*_{\epsilon_2}(\textbf{r})}e^{-i\textbf{q}\cdot\textbf{r}}$, and we recall that $\epsilon=(\epsilon_1+\epsilon_2)/2$ and $\omega=\epsilon_1-\epsilon_2$. 

The building blocks in Figs. \ref{diagrams}e and \ref{diagrams}f are calculated in the same way. For instance, the diagram \ref{diagrams}e yields the integral equation
\begin{eqnarray}
\label{BS_eq_e}
\overline{\psi_{\epsilon_1}(\textbf{r})\psi^*_{\epsilon_2}(\textbf{r})} |_\text{e}&=&
2\gamma^2 g\int\!\!\!\int\!\!\!\int d^3\textbf{r}^\prime d^3\textbf{r}_1d^3\textbf{r}_2
\int\!\!\!\int\dfrac{d\epsilon_3}{2\pi}\dfrac{d\epsilon_4}{2\pi}
\overline{G}^*_{\epsilon_2}(\textbf{r}^\prime,\textbf{r})
\overline{G}_{\epsilon_1-\epsilon_3+\epsilon_4}(\textbf{r}^\prime,\textbf{r}_2)
\overline{G}_{\epsilon_1-\epsilon_3+\epsilon_4}(\textbf{r}_2,\textbf{r}_1)
\overline{G}_{\epsilon_1}(\textbf{r}_1,\textbf{r}_2)
\overline{G}_{\epsilon_1}(\textbf{r}_2,\textbf{r}) \times\nonumber\\
&& \times\overline{\psi_{\epsilon_3}(\textbf{r})\psi^*_{\epsilon_4}(\textbf{r}_1)}\times
\overline{\psi_{\epsilon_1-\epsilon_3+\epsilon_4}(\textbf{r})\psi^*_{\epsilon_2}(\textbf{r}^\prime)},
\end{eqnarray}
where we used Eq. (\ref{correlation}) to account for the additional impurity line. Once again, the integral equation for the building block in Fig. \ref{diagrams}f follows from Eq. (\ref{BS_eq_e}) after the substitutions $\overline{G}\leftrightarrow\overline{G}^*$, $\epsilon_1\leftrightarrow\epsilon_2$ and $\epsilon_3\leftrightarrow\epsilon_4$. Summation of the building blocks e and f leads to (see Appendix \ref{appendix_B})
\begin{eqnarray}
\label{BS_eq_ef}
n_\epsilon(\textbf{r},\omega)|_\text{e}+
n_\epsilon(\textbf{r},\omega)|_\text{f}&=&
4g\tau\int\!\!\!\int \dfrac{dE}{2\pi}\dfrac{d\Omega}{2\pi}
\int\dfrac{d^3\textbf{q}_1}{(2\pi)^3}
n_E(\textbf{q}_1,\Omega)
n_\epsilon(\textbf{q}-\textbf{q}_1,\omega-\Omega)\times\nonumber\\
&&\times\left[\dfrac{1+2i(\omega-\Omega)\tau}{2k\ell}+\dfrac{\ell^2}{2k\ell}(-\textbf{q}^2+\textbf{q}\cdot\textbf{q}_1-\dfrac{\textbf{q}_1^2}{3})\right].
\end{eqnarray}
In order to complete the diagrammatic treatment, one finally needs to include the building block in Fig. \ref{diagrams}b, which describes a usual linear scattering process. This type of process may of course also occur in a nonlinear medium (see Fig. \ref{general_diagram}). When this single building block is taken into account, we saw in Sec. \ref{section4} that it generates a ladder diagram described by the Bethe-Salpeter  equation Eq. (\ref{BS_eq}). In the nonlinear case, it generates a series of diagrams for which both scattering sequences associated with $\psi_{\epsilon_1}$ and $\psi^*_{\epsilon_2}$ emerge at point $\textbf{r}$ from a linear scattering process. This series of diagrams contributes of an amount $\overline{\psi_{\epsilon_1}(\textbf{r})\psi^*_{\epsilon_2}(\textbf{r})}|_\text{b}$ to the total density, and obeys 
\begin{equation}
\label{BS_eq_2}
\overline{\psi_{\epsilon_1}(\textbf{r})\psi^*_{\epsilon_2}(\textbf{r})}|_\text{b}=
\overline{\psi}_{\epsilon_1}(\textbf{r})\overline{\psi}^*_{\epsilon_2}(\textbf{r})+\gamma\int d^3\textbf{r}^\prime\,\overline{G}_{\epsilon_1}(\textbf{r}^\prime,\textbf{r})\overline{G}^*_{\epsilon_1}(\textbf{r}^\prime,\textbf{r})\overline{\psi_{\epsilon_1}(\textbf{r}^\prime)\psi^*_{\epsilon_2}(\textbf{r}^\prime)}.
\end{equation}
The simplification of Eq. (\ref{BS_eq_2}) follows exactly the same lines as for the Bethe-Salpeter equation (see Appendix \ref{appendix_A}). By analogy with Eq. (\ref{BS_eq_simplified}), we have
\begin{equation}
\label{BS_eq_b}
n_\epsilon(\textbf{q},\omega)|_\text{b}=
S^\prime_\epsilon(\textbf{q},\omega)+
(1+i\omega\tau-D_\epsilon\tau\textbf{q}^2)\,
n_\epsilon(\textbf{q},\omega),
\end{equation}
where we have defined the source term $S^\prime_\epsilon(\textbf{q},\omega)=\int d^3\textbf{r}\,\overline{\psi}_{\epsilon_1}(\textbf{r})\overline{\psi}^*_{\epsilon_2}(\textbf{r})e^{-i\textbf{q}\cdot\textbf{r}}$. The prime symbol in this definition signals that in the nonlinear case, the source term has not the same value as in the linear case. The expression of $S^\prime_\epsilon(\textbf{q},\omega)$ will be given in the next section.

\subsection{Combination of nonlinear diagrams}

In order to complete the derivation of the Fokker-Planck equation (\ref{FP_eq}), we have to close Eqs. (\ref{BS_eq_cd}), (\ref{BS_eq_ef}), and (\ref{BS_eq_b}). This is achieved by writing
\begin{equation}
\label{total_density}
n_\epsilon(\textbf{q},\omega)=
n_\epsilon(\textbf{q},\omega)|_\text{b}+
n_\epsilon(\textbf{q},\omega)|_\text{c}+
n_\epsilon(\textbf{q},\omega)|_\text{d}+
n_\epsilon(\textbf{q},\omega)|_\text{e}+
n_\epsilon(\textbf{q},\omega)|_\text{f}.
\end{equation}
Using Eq. (\ref{BS_eq_cd}), (\ref{BS_eq_ef}) and (\ref{BS_eq_b}), we rewrite Eq. (\ref{total_density}) as
\begin{eqnarray}
\label{FP_eq_Fourier}
-S^\prime_\epsilon(\textbf{q},\omega)&=&
(i\omega\tau-D_\epsilon\tau\textbf{q}^2)n_\epsilon(\textbf{q},\omega)+\nonumber\\
&&+4g\tau\int\!\!\!\int\dfrac{dE}{2\pi}\dfrac{d\Omega}{2\pi}\int\dfrac{d^3\textbf{q}_1}{(2\pi)^3}
n_E(\textbf{q}_1,\omega)n_\epsilon(\textbf{q}-\textbf{q}_1,\omega-\Omega)\left[\dfrac{i(\omega-\Omega)\tau}{2k\ell}-\dfrac{\ell^2}{2k\ell}\dfrac{\textbf{q}_1^2}{3}\right]+\nonumber\\
&&+2ig\tau\int\!\!\!\int \dfrac{dE}{2\pi}\dfrac{d\Omega}{2\pi}
\int\dfrac{d^3\textbf{q}_1}{(2\pi)^3}
\Omega\, n_E(\textbf{q}_1,\Omega)
\partial_\epsilon n_\epsilon(\textbf{q}-\textbf{q}_1,\omega-\Omega)
\end{eqnarray}
We now divide Eq. (\ref{FP_eq_Fourier}) by $\tau$, and take the inverse Fourier transform with respect to $\textbf{q}$ and $\omega$. We obtain
\begin{eqnarray}
\label{FP_eq_real}
\dfrac{S^\prime_\epsilon(\textbf{r},t)}{\tau}&=&
-D_\epsilon\boldsymbol{\nabla}^2n_\epsilon(\textbf{r},t)+
\left[1+\dfrac{2g\tau}{k\ell}\int\dfrac{dE}{2\pi}n_E(\textbf{r},t)\right]\partial_t n_\epsilon(\textbf{r},t)-\nonumber\\
&&-\dfrac{2g\ell^2}{3k\ell}\left[\boldsymbol{\nabla}^2\int\dfrac{dE}{2\pi}n_E(\textbf{r},t)\right]n_\epsilon(\textbf{r},t)+2g\left[\partial_t\int\dfrac{dE}{2\pi}n_E(\textbf{r},t)\right]\partial_\epsilon n_\epsilon(\textbf{r},t).
\end{eqnarray}
From Eq. (\ref{FP_eq_real}), we see the emergence of the ``potential" $\theta(\textbf{r},t)=2g\int(dE/2\pi)n_E(\textbf{r},t)=2gn(\textbf{r},t)$. Since $2 \tau g n/(k\ell)\sim2\tau g n/(\epsilon\tau)\ll1$, we can divide both sides of Eq. (\ref{FP_eq_real}) by $1+2 \tau g n/(k\ell)$ and expand the result for small nonlinearity. Using  $\ell^2/(3k\ell)=D_\epsilon/(2\epsilon)$ in 3D, we can rewrite Eq. (\ref{FP_eq_real}) as
\begin{eqnarray}
\label{FP_eq_smalltheta}
\dfrac{S^\prime_\epsilon(\textbf{r},t)}{\tau}=
\left[\partial_t -D_{\epsilon}\boldsymbol{\nabla}^2+\partial_t \theta(\textbf{r},t)\partial_\epsilon\right]n_\epsilon(\textbf{r},t)+\dfrac{D_\epsilon}{2\epsilon}\left[\theta(\textbf{r},t)\boldsymbol{\nabla}^2n_\epsilon(\textbf{r},t)-n_\epsilon(\textbf{r},t)\boldsymbol{\nabla}^2\theta(\textbf{r},t)\right].
\end{eqnarray}
In order to complete the proof, we have also to evaluate the source term which appears in the left-hand side of Eq. (\ref{FP_eq_smalltheta}). Although this term describes a propagation free of linear scattering processes, it is likely to be affected by the nonlinearity \cite{Wellens09}. The result (\ref{source_term}) obtained in the Sec. \ref{section4} should therefore be reconsidered here. The calculation of $S^\prime(\textbf{r},t)$ is achieved by summing a series of diagrams of the type in Figs. 1c and 1d, but where the correlation functions $\overline{\psi\psi^*}$ of the field at points $\textbf{r}^\prime$ and $\textbf{r}_1$ are replaced by the correlation function $\overline{\phi\phi^*}$ of the initial wave function of the wave packet, and where an arbitrary number of nonlinear scattering processes may occur between the points $\textbf{r}^\prime$ and $\textbf{r}$. This calculation is lengthy and is reported to Appendix \ref{appendix_C} for clarity. It yields:
\begin{equation}
\label{source_term_nonlinear}
S^\prime(\epsilon,\textbf{r},t)=\tau\delta(t)n_{\epsilon-\theta}(\textbf{r},t=0),
\end{equation}
which is the same expression as in the linear case but with the total energy $\epsilon$ replaced by the kinetic energy $\epsilon-\theta$. The density $n_{\epsilon-\theta}(\textbf{r},t=0)$ of the wave packet at the initial time appearing in Eq. (\ref{source_term_nonlinear}) is still given by Eq. (\ref{source_term}), but the spectral function $A$ must now be evaluated at the energy $\epsilon-\theta$.
We now combine Eqs. (\ref{FP_eq_smalltheta}) and (\ref{source_term_nonlinear}), and slightly rearrange the terms in the right-hand side. This yields
\begin{eqnarray}
\label{FP_eq_smalltheta2}
\delta(t)n_{\epsilon-\theta}(\textbf{r},t=0)=
\left[\partial_t+\partial_t \theta\partial_\epsilon\right]n_\epsilon(\textbf{r},t)-
\boldsymbol{\nabla}\left[\left(D_\epsilon-\dfrac{D_\epsilon\theta}{2\epsilon}\right)\boldsymbol{\nabla} n_\epsilon(\textbf{r},t)\right]-
\boldsymbol{\nabla}\left[\dfrac{D_\epsilon}{2\epsilon}(\boldsymbol{\nabla} \theta)n_\epsilon(\textbf{r},t)\right],
\end{eqnarray}
If we notice that $D_\epsilon-D_\epsilon\theta/(2\epsilon)\simeq D_{\epsilon-\theta}$ and $D_\epsilon/(2\epsilon)=(d/2-1)D_\epsilon/\epsilon\equiv\eta_\epsilon$ in 3D, we see that Eq. (\ref{FP_eq_smalltheta2}) is nothing but the Fokker-Planck equation (\ref{FP_eq}) in the limit $g\tau n(\textbf{r},t)\ll1$, which completes the microscopic derivation.

\section{Diagrammatic treatment in 1D and 2D}
\label{section6}

In this section we briefly discuss how the above derivation is modified in 1D and 2D. The building blocks to consider remain evidently the same as in the three-dimensional case (see Figs. \ref{diagrams}b, \ref{diagrams}c, \ref{diagrams}d, \ref{diagrams}e and \ref{diagrams}f), but their calculation leads to different expressions. The two-dimensional case was already sketched in \cite{Schwiete10}: the last two terms in the right-hand side in Eq. (\ref{FP_eq_smalltheta2}) are then replaced by
\begin{eqnarray}
\label{2dcase}
\dfrac{D_\epsilon}{\epsilon}\left[\boldsymbol{\nabla}\theta(\textbf{r},t)\boldsymbol{\nabla}n_\epsilon(\textbf{r},t)+\theta(\textbf{r},t)\boldsymbol{\nabla}^2n_\epsilon(\textbf{r},t)\right],
\end{eqnarray}
which yields the Fokker-Planck equation (\ref{FP_eq}) because in 2D, $\eta_\epsilon=0$, and $D_{\epsilon-\theta}= D_\epsilon-D_\epsilon\theta/\epsilon $. Finally, for completeness we also analyze the one-dimensional case, although it is less relevant due to the existence of strong localization effects that take place at time scales of the order of the scattering time \cite{Abrahams79}. In 1D, the last two terms in the right-hand side in Eq. (\ref{FP_eq_smalltheta2}) are replaced by:
\begin{eqnarray}
\label{1dcase}
\dfrac{D_\epsilon}{\epsilon}
\left[2\boldsymbol{\nabla}\theta(\textbf{r},t)\boldsymbol{\nabla}n_\epsilon(\textbf{r},t)+\dfrac{3}{2}\theta(\textbf{r},t)\boldsymbol{\nabla}^2n_\epsilon(\textbf{r},t)+\dfrac{1}{2}n_\epsilon(\textbf{r},t)\boldsymbol{\nabla}^2\theta(\textbf{r},t)\right].
\end{eqnarray}
In 1D, $\eta_\epsilon=-D_\epsilon/(2\epsilon)$, and $D_{\epsilon-\theta} \simeq D_\epsilon-3D_\epsilon\theta/(2\epsilon)$ in the limit $\tau g n(\textbf{r},t)\ll1$, such that Eq. (\ref{1dcase}) leads once again to the Fokker-Planck equation.

\section{Heuristic derivation of the Fokker-Planck equation}
\label{section7}

In the microscopic derivation presented in the previous sections, the ``nonlinear potential" $\theta(\textbf{r},t)=2 g n(\textbf{r},t)$ emerges naturally. Note that $\theta$ differs by a factor 2 from the term $g n(\textbf{r},t)$ added to the potential $V$ in the Gross-Pitaevskii equation (\ref{GPE}). Apart from that, it turns out that the complicated problem of wave packet propagation in a nonlinear disordered medium can be replaced by the more simple one of a collection of independent, classical particles subject to an external force, and moving in a disordered environment. The purpose of this section is to confirm this picture, and to take advantage of it to present a very simple way of deriving the Fokker-Plank equation (\ref{FP_eq}). 

For a system of independent particles subject to the ``external" force $\textbf{F}=-\boldsymbol{\nabla}\theta$, the transport properties are contained in the one-particle density function $f(\textbf{r},\textbf{p},t)$, which gives the probability density of finding a particle at position $\textbf{r}$ with momentum $\textbf{p}$ and kinetic energy $\epsilon_0=\textbf{p}^2/(2m)$. Since we are considering a disordered medium with point-like scatterers, the time evolution of $f$ is governed by the Boltzmann equation 
\begin{equation}
\label{Boltzmann}
\left[\partial_t+\dfrac{\textbf{p}}{m}\cdot\boldsymbol{\nabla}+\textbf{F}\cdot\boldsymbol{\nabla}_\textbf{p}\right]f(\textbf{r},\textbf{p},t)=
\dfrac{f(\textbf{r},\textbf{p},t)-f(\textbf{r},p,t)}{\tau}.
\end{equation}
Eq. (\ref{Boltzmann}) expresses the fact that the rate of change in $f$ in phase-space $(\textbf{r},\textbf{p})$ is due to the scattering from impurities. The density $n_{\epsilon_0}(\textbf{r},t)$ introduced in the previous sections is related to $f$ through the simple relation $n_{\epsilon_0}(\textbf{r},t)=\nu\, f(\textbf{r},p,t)$, with $\nu$ the density of states per unit volume. We stress at this point that $\epsilon_0$ does designates the \emph{kinetic} energy, which should not be confused with the \emph{total} energy $\epsilon$ of Secs. \ref{section5} and \ref{section6}. $f(\textbf{r},p,t)=\int d \boldsymbol{\Omega}_p/(4\pi) f(\textbf{r},\textbf{p},t)$ is the average of $f(\textbf{r},\textbf{p},t)$ over all directions $\boldsymbol{\Omega}_p$ of $\textbf{p}$. In order to find a transport equation for $f(\textbf{r},p,t)$, we make use of the diffusion approximation, which here consists in expanding the density function into spherical harmonics, keeping only the lowest two harmonics \cite{Ishimaru}:
\begin{equation}
\label{hamonics}
 f(\textbf{r},\textbf{p},t)\simeq  f(\textbf{r},p,t)+d\,\boldsymbol{\Omega}_p\cdot \textbf{f}_1(\textbf{r},p,t),
\end{equation}
 where $\textbf{f}_1(\textbf{r},p,t)=\int d \boldsymbol{\Omega}_p/(4\pi)\boldsymbol{\Omega}_p f(\textbf{r},\textbf{p},t)$, and $d=1,2,3$. Substituting this expansion into Eq. (\ref{Boltzmann}), we obtain an equation, which, after integration over angles, yields 
\begin{equation}
\label{Bol1}
\partial_tf(\textbf{r},p,t)+\left[\dfrac{p}{md}\boldsymbol{\nabla}+ \dfrac{p}{md}\textbf{F}\partial_{\epsilon_0}+\left(1-\dfrac{1}{d}\right)\dfrac{\textbf{F}}{p}\right] \cdot\textbf{f}_1(\textbf{r},p,t)=0,
\end{equation}
and, after multiplication by $\boldsymbol{\Omega}_p$ and integration over angles,
\begin{equation}
\label{Bol2}
\textbf{f}_1(\textbf{r},p,t)=-\dfrac{p\tau}{m}\left(\boldsymbol{\nabla}+\textbf{F}\partial_{\epsilon_0}\right) f(\textbf{r},p,t).
\end{equation}
Finally, combining Eqs. (\ref{Bol1}) and (\ref{Bol2}), using the relation $n_{\epsilon_0}(\textbf{r},t)=\nu\, f(\textbf{r},p,t)$ and writing the initial value $n_{\epsilon_0}(\textbf{r},t=0)$ explicitly in form of a source term in the right-hand side,  we obtain
\begin{equation}
\label{FP_eq_epsilon}
\partial_tn_{\epsilon_0}(\textbf{r},t)-
\left[\boldsymbol{\nabla}+\textbf{F}\partial_{\epsilon_0}\right]
D_{\epsilon_0}
\left[\boldsymbol{\nabla}+\textbf{F}\partial_{\epsilon_0}-(\eta_{\epsilon_0}/D_{\epsilon_0})\textbf{F}\right]n_{\epsilon_0}(\textbf{r},t)=\delta(t)n_{\epsilon_0}(\textbf{r},t=0),
\end{equation}
where $\eta_{\epsilon_0}=(d/2-1)D_{\epsilon_0}/\epsilon_0$. Eq. (\ref{FP_eq_epsilon}) was obtained in \cite{Schwiete10} in the particular two-dimensional case, for which the parameter $\eta_{\epsilon_0}=0$. Eq. (\ref{FP_eq_epsilon}) is the generalization of this result to any dimension. The Fokker-Planck equation (\ref{FP_eq}) follows straightforwardly after the transformation ${\epsilon_0}\rightarrow\epsilon-\theta$ (with $\epsilon$ the total energy), 
in agreement with the diagrammatic treatment. Note that in this section we only made use of two approximations. First, we resorted to the diffusion approximation, which allowed us  to use the expansion (\ref{hamonics}). Second, when writing the Boltzmann equation (\ref{Boltzmann}), we assumed that the scattering time was unaffected by the nonlinearity, which is valid as long as condition (\ref{cond_tau}) is fulfilled. Unlike in the previous sections however, we did not use the condition $\tau g n(\textbf{r},t)\ll1$. This means that Eq. (\ref{FP_eq_epsilon}), as well as the Fokker-Planck (\ref{FP_eq}), are not restricted to this limit.

\section{Concluding remarks}
\label{section8}

In this paper, we have formulated a transport theory of wave packet propagation in a nonlinear, weakly disordered medium, through the microscopic derivation of the Fokker-Planck equation (\ref{FP_eq}). The latter describes transport in media of any dimension, and thereby generalizes the equation introduced in \cite{Schwiete10}. Our derivation is based on a diagrammatic formalism, under the assumption of weak disorder, and in the framework of the diffusion approximation. Within this approach, we have identified all the diagrams required to ensure conservation of the total density $n(\textbf{r},t)$, and we have derived the Fokker-Planck equation in the limit of small nonlinearity ($\tau g n\ll1$). As compared to the linear case, the main difficulty of the nonlinear diagrammatic procedure presented in this paper is that nonlinear terms always show up in powers of the parameter $\tau gn/(\epsilon\tau)$. This can be seen, for instance, in Eq. (\ref{FP_eq_smalltheta}), where the nonlinear terms $\partial_t\theta\partial_\epsilon$, $(D_\epsilon/\epsilon)\theta\boldsymbol{\nabla}^2$ and $(D_\epsilon/\epsilon)\boldsymbol{\nabla}^2 \theta $ in the right-hand side are all of order $\tau gn/(\epsilon\tau)$. The perturbation theory has therefore to be carried out carefully, keeping terms up to the first order in  $\tau g n/(\epsilon\tau)$. Note that the fact that nonlinear corrections to the diffusive transport depend on the single parameter $\tau gn/(\epsilon\tau)$ means that it is not easily possible to decouple the nonlinearity from the strength of the disorder within a perturbative approach. This effect was already pointed out in the case of the three-dimensional expansion of a Bose-Einstein condensate in a disordered medium, in the presence of very weak interactions \cite{Cherroret09}.

We have also shown that the physical problem of wave packet expansion in a random potential can be conveniently described by a heuristic approach, in which the wave packet is replaced by a collection of independent, classical particles subject to an external, effective potential. This approach, based on the Boltzmann transport equation, allowed us to recover the Fokker-Planck equation, thereby validating the diagrammatic procedure. Unlike the latter however, the derivation from the Boltzmann equation is non-perturbative, and only relies on the diffusion approximation and the fact that the scattering time is not affected by the nonlinearity. This means that the Fokker-Planck equation (\ref{FP_eq}) remains valid for possibly large values of $\tau g n$, on condition that $\tau g n$ does not exceed $\sqrt{\epsilon\tau}$.

Let us conclude our study by a discussion on the practical validity of the Fokker-Planck equation in the time domain. In deriving Eq. (\ref{FP_eq}), we have assumed that no coherent effects of localization type take place in the disordered medium. This assumption limits the range of validity of the description at long times, when the mean radius of the wave packet reaches the localization length $\xi_{\epsilon_0}$, where $\epsilon_0$ is the typical kinetic energy of the wave packet. For the particular case of a Bose-Einstein condensate expanding in a random potential after release from an optical
trap, $\epsilon_0$ is of the order of the chemical potential $\mu$ \cite{Shapiro07}. In 1D and 2D, we can estimate the time $t_\text{loc}$ after which the approach breaks down due to localization effects, by requiring that $\langle\textbf{r}^2\rangle\sim\xi_{\mu}^2$, where $\langle\textbf{r}^2\rangle\sim D_\mu t_\text{loc}$. In 1D, this yields $t_\text{loc}\sim\tau$, which means that the diffusive approach breaks down at short times, and thus limits the relevance of the present work in this case. In 2D, $t_\text{loc}\sim\tau\exp(\mu\tau)\gg\tau$, such that a broad diffusion regime exists where the Fokker-Planck equation applies. In 3D, the situation is slightly more complicated since there exists a critical energy, separating diffusive and localized states. It was shown recently that in this case, the diffusive approach, and consequently the Fokker-Planck equation (\ref{FP_eq}), is valid, at distance $r$ from the origin, as long as time does not exceed $t_\text{loc}\sim\tau(r/\ell)^{8/3}(\epsilon\tau)^2\gg\tau$ \cite{Cherroret09}.

The transport equation (\ref{FP_eq}) is a flexible theoretical tool, and we think it could be used as a basis for future work on time-dependent problems in disordered and nonlinear media, for instance in the context of Bose-Einstein condensates expanding in random potentials. Although strictly speaking we have only addressed transport in the infinite medium, Eq. (\ref{FP_eq}) holds in finite media as well, provided it is supplemented by adequate boundary conditions \cite{Pine91}. Since the nonlinear effects only weakly affect transport at the scale of one mean free path, we expect the boundary conditions used in the linear problem to remain approximately valid in nonlinear media.

\section{Acknowledgments}

We thank Denis Basko for useful discussions on the Boltzmann equation, and Andreas Buchleitner for his comments on the manuscript. N.C. acknowledges financial support from the Alexander von Humboldt foundation.

\appendix
\section{}
\label{appendix_A}

In this appendix we present a derivation of Eq. (\ref{BS_eq_simplified}), starting from the Bethe-Salpeter equation (\ref{BS_eq}). For this purpose, it is convenient to work in Fourier space. The Fourier transform of the second term in the right-hand side of Eq. (\ref{BS_eq}) is given by
\begin{equation}
\label{A1}
\gamma\int \dfrac{d^3\textbf{k}}{(2\pi)^3}\overline{G}_{\epsilon_1}(\textbf{q})\overline{G}^*_{\epsilon_2}(\textbf{k}+\textbf{q})n_\epsilon(\textbf{q},\omega),
\end{equation}
where $n_\epsilon(\textbf{q},\omega)=\int d^3\textbf{r}\, \overline{\psi_{\epsilon_1}(\textbf{r})\psi^*_{\epsilon_2}(\textbf{r})}e^{-i\textbf{q}\cdot\textbf{r}}$, with $\epsilon=(\epsilon_1+\epsilon_2)/2$ and $\omega=\epsilon_1-\epsilon_2$. We have also introduced the Fourier transform of the average amplitude Green's function \cite{AM}
\begin{equation}
\label{GF_Fourier}
\overline{G}_\epsilon(\textbf{q})=\dfrac{1}{\epsilon-\epsilon_q+i/(2\tau)},
\end{equation}
with $\epsilon_q=\textbf{q}^2/(2m)$. In the diffusion regime, $|\textbf{q}\ell |\ll1$, and we can expand the Green's function at energy $\epsilon_2$ according to
\begin{equation}
\label{GF_expansion}
\overline{G}^*_{\epsilon_2}(\textbf{k}+\textbf{q})\simeq\overline{G}^*_{\epsilon_2}(\textbf{k})+[(\textbf{v}\cdot\textbf{q})+\epsilon_q]\overline{G}^*_{\epsilon_2}(\textbf{k})^2+(\textbf{v}\cdot\textbf{q})\overline{G}^*_{\epsilon_2}(\textbf{k})^3,
\end{equation}
where $\textbf{v}=\textbf{q}/m$. Inserting this expansion into Eq. (\ref{A1}), and performing integrations over $\textbf{k}$, we obtain
\begin{equation}
\label{A2}
\gamma\int \dfrac{d^3\textbf{k}}{(2\pi)^3}\overline{G}_{\epsilon_1}(\textbf{q})\overline{G}^*_{\epsilon_2}(\textbf{k}+\textbf{q})n_\epsilon(\textbf{q},\omega)\simeq
(1+i\omega\tau-\dfrac{v^2\tau^2}{3}\textbf{q}^2)n_\epsilon(\textbf{q},\omega).
\end{equation}
This result, combined with Eq. (\ref{BS_eq}), yields Eq. (\ref{BS_eq_simplified}), with $D_\epsilon=v^2\tau/3$. In 1D and 2D, the derivation follows exactly the same lines, the only difference being the replacement of the diffusion coefficient by $D_\epsilon=v^2\tau/d$. Note that in deriving Eq. (\ref{A2}), we neglected all terms of order $1/(\epsilon\tau)$ or higher (limit of weak disorder), and kept only terms linear in $\omega\tau$ and $\textbf{q}^2\ell^2$ (diffusion approximation).

\section{}
\label{appendix_C}

\begin{figure}[h]
\includegraphics[width=12cm]{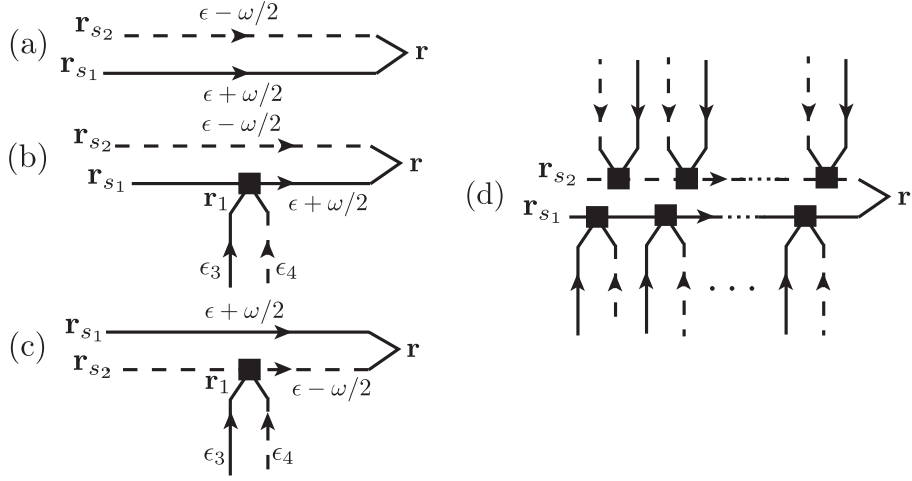}
\caption{\label{diagrams_source} 
(a) Diagram describing the source term in the absence of nonlinearity. $\textbf{r}_{s_1}$ and $\textbf{r}_{s_2}$ are two points of the wave packet at initial time. (b)  and (c) Leading-order diagrams contributing to the source term in the presence of nonlinearity, in the limit $\tau g n\ll1$. (d) Typical diagram contributing to the source term, allowing to account for large values of the nonlinearity.}  
\end{figure}

In this appendix we calculate the ``source terms" $S_ \epsilon(\textbf{r},t)$ and $S^\prime_ \epsilon(\textbf{r},t)$ in the diffusion and Fokker-Planck equations (\ref{diff_eq_2}) and (\ref{FP_eq_smalltheta}), respectively. Both are defined as
\begin{equation}
\label{C1}
S^{(\prime)}_\epsilon(\textbf{r},t)\equiv
\int \dfrac{d\omega}{2\pi}\,\overline{\psi}_{\epsilon+\omega/2}(\textbf{r})\overline{\psi}^*_{\epsilon-\omega/2}(\textbf{r})e^{-i\omega t}.
\end{equation}
To begin with, we calculate $S_\epsilon$, \emph{i.e.} the source term in the absence of nonlinearity. To do so,  it is convenient to relate the wave function $\psi$ at point $\textbf{r}$ to the wave function $\phi$ of the wave packet at the initial time through
\begin{equation}
\label{C2}
\psi_\epsilon(\textbf{r})=\int d^3 \textbf{r}_{s} G_{\epsilon}(\textbf{r}_s,\textbf{r})\phi(\textbf{r}_s).
\end{equation}
Substituting Eq. (\ref{C2}) into Eq. (\ref{C1}), we obtain
\begin{equation}
\label{source_def}
S_\epsilon(\textbf{r},t)=\int \dfrac{d\omega}{2\pi} e^{-i\omega t}\int \!\!\! \int d^3 \textbf{r}_{s_1} d^3 \textbf{r}_{s_2} \overline{G}_{\epsilon+\omega/2}(\textbf{r}_{s_1},\textbf{r})\overline{G}^*_{\epsilon-\omega/2}(\textbf{r}_{s_2},\textbf{r})\phi(\textbf{r}_{s_1})\phi^*(\textbf{r}_{s_2}).
\end{equation}
The diagrammatic representation of Eq. (\ref{source_def}) is shown in Fig. \ref{diagrams_source}a. Introducing the Fourier transforms $\overline{G}(\textbf{q})=\int d^3\textbf{r}\,\overline{G}(\textbf{r})e^{-i \textbf{q}\cdot\textbf{r}}$ of the average amplitude Green's functions appearing in Eq. (\ref{source_def}), we obtain
\begin{equation}
\label{source_def_Fourier}
S_ \epsilon(\textbf{r},t)=\int \dfrac{d\omega}{2\pi} e^{-i\omega t} \int \!\!\! \int \frac{d^3 \textbf{q}}{(2\pi)^3} \frac{d^3 \textbf{Q}}{(2\pi)^3}\,\overline{G}_{\epsilon+\omega/2}\left(\textbf{Q}+\textbf{q}/2\right)\overline{G}^*_{\epsilon-\omega/2}\left(\textbf{Q}-\textbf{q}/2\right)\phi\left(\textbf{Q}+\textbf{q}/2\right)\phi^*\left(\textbf{Q}-\textbf{q}/2\right).
\end{equation}
We now perform the integral over $\omega$. Using $|\textbf{q}|\ll |\textbf{Q}|$, $\omega\ll\epsilon$ (diffusion approximation), and $\epsilon\tau\gg1$ (limit of weak disorder, we find:
\begin{equation}
\int \dfrac{d\omega}{2\pi} e^{-i\omega t}\,
\overline{G}_{\epsilon+\omega/2}(\textbf{Q}+\textbf{q}/2)\overline{G}^*_{\epsilon-\omega/2}(\textbf{Q}-\textbf{q}/2)\simeq 2\pi\tau\delta(t)A(\epsilon,\textbf{Q}),
\end{equation}
where we have introduced the spectral function $A(\epsilon,\textbf{Q})\equiv-\text{Im}\overline{G}_\epsilon(\textbf{Q})/\pi$. This yields:
\begin{equation}
\label{Source_appendix_A}
S(\epsilon,\textbf{r},t)=\tau\delta(t)\int \!\!\! \int \frac{d^3 \textbf{q}}{(2\pi)^3} \frac{d^3 \textbf{Q}}{(2\pi)^3}2\pi A(\epsilon,\textbf{Q})\phi\left(\textbf{Q}+\textbf{q}/2\right)\phi^*\left(\textbf{Q}-\textbf{q}/2\right)e^{-i\textbf{q}\cdot\textbf{r}}.
\end{equation}
Eq. (\ref{Source_appendix_A}) is the source term that appears in the diffusion equation (\ref{diff_eq_2}). It can be readily related to the wave packet's density at the initial time if we note that $n_\epsilon(\textbf{r},t=0)=\int(d\omega/2\pi)\overline{\psi_{\epsilon+\omega/2}(\textbf{r})\psi^*_{\epsilon-\omega/2}(\textbf{r})}\simeq\int(d\omega/2\pi)\overline{\psi}_{\epsilon+\omega/2}(\textbf{r})\overline{\psi}^*_{\epsilon-\omega/2}(\textbf{r})$. Following exactly the same reasoning as above, we find
\begin{equation}
n_\epsilon(\textbf{r},t=0)=\int \!\!\! \int \frac{d^3 \textbf{q}}{(2\pi)^3} \frac{d^3 \textbf{Q}}{(2\pi)^3}2\pi A(\epsilon,\textbf{Q})\phi\left(\textbf{Q}+\textbf{q}/2\right)\phi^*\left(\textbf{Q}-\textbf{q}/2\right)e^{-i\textbf{q}\cdot\textbf{r}},
\end{equation}
which yields immediately Eq. (\ref{source_term_2}) of the main text.

We now consider the source term $S^\prime_\epsilon$, which is affected by the nonlinearity. In comparison with the linear case where only the diagram in Fig. \ref{diagrams_source}a is relevant, 
now also the nonlinear diagrams in Figs. \ref{diagrams_source}b and \ref{diagrams_source}c contribute. For the sake of consistency with Sec. \ref{section5},  we assume that $\tau gn(\textbf{r},t)\ll1$, namely we neglect diagrams for which more than one nonlinear scattering event occurs in a row. The final result that we obtain is however valid beyond this limit, as we discuss later on. In the nonlinear case, the source term is given by
\begin{equation}
\label{SumS}
S^\prime_\epsilon(\textbf{r},t)=
S_\epsilon(\textbf{r},t)+
S^{(b)}_\epsilon(\textbf{r},t)+
S^{(c)}_\epsilon(\textbf{r},t),
\end{equation}
where the two terms with the superscripts (b) and (c) refer to the diagrams in Figs. \ref{diagrams_source}b and \ref{diagrams_source}c, respectively. Within the diffusion approximation where $\omega\tau,(\epsilon_3-\epsilon_4)\tau\ll1$, $S_\epsilon^{(b)}$ is for instance given by
\begin{eqnarray}
\label{Sb}
S^{(b)}_\epsilon(\textbf{r},t)&\simeq&
2g\int\dfrac{d\omega}{2\pi}e^{-i\omega t}
\int\!\!\!\int\!\!\!\int d^3\textbf{r}_{s_1}d^3\textbf{r}_{s_2}d^3\textbf{r}_1
\int\!\!\!\int\dfrac{d\epsilon_3}{2\pi}\dfrac{d\epsilon_4}{2\pi}
\overline{G}_{\epsilon}(\textbf{r}_{s_1},\textbf{r}_1)
\overline{G}_{\epsilon}(\textbf{r}_1,\textbf{r})
\overline{G}^*_{\epsilon}(\textbf{r}_{s_2},\textbf{r})\times\\\nonumber
&\times&\phi(\textbf{r}_{s_1})\phi^*(\textbf{r}_{s_2})\times
\overline{\psi_{\epsilon_3}(\textbf{r}_1)\psi^*_{\epsilon_4}(\textbf{r}_2)}.
\end{eqnarray}
We now approximate $\overline{\psi_{\epsilon_3}(\textbf{r}_1)\psi^*_{\epsilon_4}(\textbf{r}_1)}\simeq \overline{\psi_{\epsilon_3}(\textbf{r})\psi^*_{\epsilon_4}(\textbf{r})}$, and perform the integrals over $\epsilon_3$, $\epsilon_4$, $\omega$ and $\textbf{r}_1$, by using the following result, valid in the limit of weak disorder:
\begin{equation}
\int d^3\textbf{r}_1 \overline{G}_{\epsilon}(\textbf{r}_{s_1},\textbf{r}_1)\overline{G}_{\epsilon}(\textbf{r}_1,\textbf{r})\simeq
-i\tau\frac{|\textbf{r}-\textbf{r}_{s_1}|}{\ell} \overline{G}_{\epsilon}(\textbf{r}_{s_1},\textbf{r}).
\end{equation}
Eq. (\ref{Sb}) then becomes
\begin{eqnarray}
\label{Sbfinal}
S^{(b)}_\epsilon(\textbf{r},t)&=&
-\delta(t)
\int\!\!\!\int d^3\textbf{r}_{s_1}d^3\textbf{r}_{s_2}
\dfrac{2i\tau g|\textbf{r}-\textbf{r}_{s_1}|}{\ell}n(\textbf{r},t=0)
\overline{G}_{\epsilon}(\textbf{r}_{s_1},\textbf{r})
\overline{G}^*_{\epsilon}(\textbf{r}_{s_2},\textbf{r})
\phi^*(\textbf{r}_{s_1})\phi(\textbf{r}_{s_2}).
\end{eqnarray}
The calculation of $S^{(c)}$ is identical. It leads to the same result as Eq. (\ref{Sbfinal}), but with $|\textbf{r}-\textbf{r}_{s_1}|$ replaced by $-|\textbf{r}-\textbf{r}_{s_2}|$. Then, with Eq. (\ref{source_def}), Eq. (\ref{SumS}) yields
\begin{equation}
\label{S_first}
S^\prime_\epsilon(\textbf{r},t)=
\delta(t)
\int \!\!\! \int d^3\textbf{r}_{s_1} d^3\textbf{r}_{s_2}
\exp\left(-\lambda i|\textbf{r}-\textbf{r}_{s_1}|\right)
\exp\left(\lambda i |\textbf{r}-\textbf{r}_{s_2}|\right)
\overline{G}_{\epsilon}(\textbf{r}_{s_1},\textbf{r})
\overline{G}^*_{\epsilon}(\textbf{r}_{s_2},\textbf{r})
\phi(\textbf{r}_{s_1})\phi^*(\textbf{r}_{s_2}),
\end{equation}
where $\lambda=2g\tau n(\textbf{r},t=0)/\ell$. Note that to obtain Eq. (\ref{S_first}), we made use of the approximation $1-2ig\tau n|\textbf{r}-\textbf{r}_{s_1}|/\ell+2ig\tau n|\textbf{r}-\textbf{r}_{s_2}|/\ell\simeq\exp(-2ig\tau n|\textbf{r}-\textbf{r}_{s_1}|/\ell+2ig\tau n|\textbf{r}-\textbf{r}_{s_2}|)$. This approximation, valid in the limit $\tau g n\ll1$, turns out to be the exact result obtained when one assumes that an \emph{arbitrary} number of nonlinear scattering events may affect the scattering paths associated with $G$ and $G^*$, \emph{i.e.}, when one evaluates all diagrams of the type of the one in Fig. \ref{diagrams_source}d. Eq. (\ref{S_first}) therefore holds also for large values of $\tau g n$. By introducing the Fourier transform of the Green's functions and the exponentials that appear in Eq. (\ref{S_first}), we rewrite $S^\prime_\epsilon$ as:
\begin{eqnarray}
\label{S_fourier}
S^\prime_\epsilon(\textbf{r},t)&=&
\delta(t)
\int \!\!\! \int \!\!\!\int \!\!\! \int 
\dfrac{d^3\textbf{q}}{(2\pi)^3}
\dfrac{d^3\textbf{Q}}{(2\pi)^3}
\dfrac{d^3\textbf{q}_1 }{(2\pi)^3}
\dfrac{d^3\textbf{q}_2}{(2\pi)^3}
\overline{G}_{\epsilon}(\textbf{Q}-\textbf{q}_1+\textbf{q}/2)
\overline{G}^*_{\epsilon}(\textbf{Q}+\textbf{q}_1-\textbf{q}/2)
\times\\\nonumber
&&\times
F_{-}(\textbf{q}_1)F_{+}(\textbf{q}_2)
\phi(\textbf{Q}+\textbf{q}/2)\phi^*(\textbf{Q}-\textbf{q}/2)e^{-i\textbf{q}\cdot\textbf{r}},
\end{eqnarray}
where $F_{\mp}(\textbf{q})=\int d^3\textbf{r}\exp(\mp i\lambda r-i\textbf{q}\cdot\textbf{r})=(2\pi)^3\time2\delta(q\pm\lambda)\delta(\cos \theta)/q^2$, with $\theta$ the angle between $\textbf{q}$ and the $z$-axis. Performing integrals over $\textbf{q}_1$ and $\textbf{q}_2$, and using the diffusion approximation $|\textbf{q}|\ll |\textbf{Q}|$, we obtain
\begin{eqnarray}
S^\prime_ \epsilon(\textbf{r},t)=
\delta(t)
\int \!\!\! \int
\dfrac{d^3\textbf{q}}{(2\pi)^3}
\dfrac{d^3\textbf{Q}}{(2\pi)^3}
|\overline{G}_{\epsilon}(Q+\lambda)|^2
\phi(\textbf{Q}+\textbf{q}/2)\phi^*(\textbf{Q}-\textbf{q}/2)e^{-i\textbf{q}\cdot\textbf{r}},
\end{eqnarray}
where $Q=|\textbf{Q}|$. By definition [see Eq. (\ref{GF_Fourier})], we have
\begin{equation}
|\overline{G}_{\epsilon}(Q+\lambda)|^2
=-2\tau \text{Im}\overline{G}_{\epsilon-\epsilon_{Q+\lambda}}(\textbf{Q})=2\pi\tau A(\epsilon-\epsilon_{Q+\lambda},\textbf{Q}),
\end{equation}
where we have used the definition of the spectral function $A$ in the last equality. The last step consists in writing $A(\epsilon-\epsilon_{Q+\lambda},\textbf{Q})\simeq A(\epsilon-\epsilon_{Q}-2\lambda\epsilon/k,\textbf{Q})=A[\epsilon-\epsilon_{Q}-2 g n(\textbf{r},t=0),\textbf{Q}]$, which is a good approximation as long as $\tau g n(\textbf{r},t)\ll \epsilon\tau$. This finally leads to the same expression (\ref{Source_appendix_A}) as in the linear case but with the energy $\epsilon$ replaced by $\epsilon-\theta$ in the spectral function, and completes the proof of Eq. (\ref{source_term_nonlinear}).

\section{}
\label{appendix_B}

In this appendix we derive Eqs. (\ref{BS_eq_cd}) and (\ref{BS_eq_ef}), starting from Eqs. (\ref{BS_eq_c}) and (\ref{BS_eq_e}), respectively. To do so, we introduce the notation $n(\textbf{q},\epsilon_1,\epsilon_2)\equiv n_\epsilon(\textbf{q},\omega)=\int d^3\textbf{r}\, \overline{\psi_{\epsilon_1}(\textbf{r})\psi^*_{\epsilon_1}(\textbf{r}})e^{-i\textbf{q}\cdot\textbf{r}}$, with $\epsilon=(\epsilon_1+\epsilon_2)/2$ and $\omega=\epsilon_1-\epsilon_2$. 

Consider first $n_\epsilon(\textbf{r},\omega)|_\text{c}$. Taking the Fourier transform of Eq. (\ref{BS_eq_c}) with respect to $\textbf{r}$, we obtain
\begin{equation}
\label{B1}
n_\epsilon(\textbf{r},\omega)|_\text{c}=2\gamma g \int\!\!\!\int\dfrac{d^3\textbf{q}_1}{(2\pi)^3}\dfrac{d^3\textbf{k}}{(2\pi)^3}\int\!\!\!\int\dfrac{d\epsilon_3}{2\pi}\dfrac{d\epsilon_4}{2\pi}\overline{G}^*_{\epsilon_2}(\textbf{k}-\textbf{q})\overline{G}_{\epsilon_1-\epsilon_3+\epsilon_4}(\textbf{k}-\textbf{q}_1)\overline{G}_{\epsilon_1}(\textbf{k})
n(\textbf{q}_1,\epsilon_3,\epsilon_4)
n(\textbf{q}-\textbf{q}_1,\epsilon_1-\epsilon_3+\epsilon_4,\epsilon_2)
\end{equation}
In the diffusion approximation, $|\textbf{q}\ell|$, $|\textbf{q}_1\ell |$, $|(\epsilon_1-\epsilon_2)\tau|\ll1$ and $|(\epsilon_3-\epsilon_4)\tau|\ll1$. The integral over $\textbf{k}$ can then be performed by expanding the Green's functions $\overline{G}^*_{\epsilon_2}$ and $\overline{G}_{\epsilon_1-\epsilon_3+\epsilon_4}$ in powers of $\textbf{q}$ and $\textbf{q}_1$, by means of Eq. (\ref{GF_expansion}). This leads to
\begin{eqnarray}
&&g\int \dfrac{d^3\textbf{k}}{(2\pi)^3}\overline{G}^*_{\epsilon_2}(\textbf{k}-\textbf{q})\overline{G}_{\epsilon_1-\epsilon_3+\epsilon_4}(\textbf{k}-\textbf{q}_1)\overline{G}_{\epsilon_1}(\textbf{k})
=\nonumber\\
&&\tau\left[-i+2(\omega-\Omega)\tau-\dfrac{1+i(\omega-\Omega)\tau}{2k\ell}+\ell^2(\textbf{q}^2-\textbf{q}\cdot\textbf{q}_1)\left(\dfrac{1}{2k\ell}+i\right)
+\dfrac{i\ell^2\textbf{q}_1^2}{3}\right],
\end{eqnarray}
where $\Omega=\epsilon_3-\epsilon_4$. As we pointed out in Sec. \ref{section5}, $n_\epsilon(\textbf{r},\omega)|_\text{d}$ can be straightforwardly deduced from $n_\epsilon(\textbf{r},\omega)|_\text{d}$ via the substitutions $\overline{G}\leftrightarrow\overline{G}^*$, $\epsilon_1\leftrightarrow\epsilon_2$ and $\epsilon_3\leftrightarrow\epsilon_4$. With this procedure, we can write
\begin{eqnarray}
\label{ncplusnd}
n_\epsilon(\textbf{r},\omega)|_\text{c}&+&n_\epsilon(\textbf{r},\omega)|_\text{d}=\nonumber\\
&&2g\tau\int\dfrac{d^3\textbf{q}_1}{(2\pi)^3}\int\!\!\!\int\dfrac{dE}{2\pi}\dfrac{d\Omega}{2\pi}
n_E(\textbf{q}_1,\Omega)n_{\epsilon-\Omega/2}(\textbf{q}-\textbf{q}_1,\omega-\Omega)\times\nonumber\\
&&\left[-i+2(\omega-\Omega)\tau-\dfrac{1+i(\omega-\Omega)\tau}{2k\ell}+\ell^2(\textbf{q}^2-\textbf{q}\cdot\textbf{q}_1)\left(\dfrac{1}{2k\ell}+i\right)
+\dfrac{i\ell^2\textbf{q}_1^2}{3}\right]+\nonumber\\
&&+2g\tau\int\dfrac{d^3\textbf{q}_1}{(2\pi)^3}\int\!\!\!\int\dfrac{dE}{2\pi}\dfrac{d\Omega}{2\pi}
n_E(\textbf{q}_1,\Omega)n_{\epsilon+\Omega/2}(\textbf{q}-\textbf{q}_1,\omega-\Omega)\times\nonumber\\
&&\left[i-2(\omega-\Omega)\tau-\dfrac{1+i(\omega-\Omega)\tau}{2k\ell}+\ell^2(\textbf{q}^2-\textbf{q}\cdot\textbf{q}_1)\left(\dfrac{1}{2k\ell}-i\right)
-\dfrac{i\ell^2\textbf{q}_1^2}{3}\right],
\end{eqnarray}
where we changed the integrations over $\epsilon_3$ and $\epsilon_4$ into integrations over $E=(\epsilon_3+\epsilon_4)/2$ and $\Omega=\epsilon_3-\epsilon_4$, such that $n(\textbf{q}_1,\epsilon_3,\epsilon_4)=n_E(\textbf{q}_1,\Omega)$, $n(\textbf{q}-\textbf{q}_1,\epsilon_1-\epsilon_3+\epsilon_4,\epsilon_2)=n_{\epsilon-\Omega/2}(\textbf{q}-\textbf{q}_1,\omega-\Omega)$ and $n(\textbf{q}-\textbf{q}_1,\epsilon_1,\epsilon_2+\epsilon_3-\epsilon_4)=n_{\epsilon+\Omega/2}(\textbf{q}-\textbf{q}_1,\omega-\Omega)$. The next step consists in expanding the densities $n_{\epsilon\pm\Omega/2}$ around $\epsilon$, making use of $\Omega\ll\epsilon$ (diffusion approximation). We thus have
\begin{equation}
\label{nepsilon_expansion}
n_{\epsilon\pm\Omega/2}\simeq
n_{\epsilon}\pm\dfrac{\Omega}{2}\partial_\epsilon n_\epsilon.
\end{equation}
Substituting this expansion into Eq. (\ref{ncplusnd}), we obtain
\begin{eqnarray}
\label{BS_eq_cd2}
n_\epsilon(\textbf{r},\omega)|_\text{c}&+&
n_\epsilon(\textbf{r},\omega)|_\text{d}=\nonumber\\
&&4g\tau\int\!\!\!\int \dfrac{dE}{2\pi}\dfrac{d\Omega}{2\pi}
\int\dfrac{d^3\textbf{q}_1}{(2\pi)^3}
n_E(\textbf{q}_1,\Omega)
n_\epsilon(\textbf{q}-\textbf{q}_1,\omega-\Omega)
\left[-\dfrac{1+i(\omega-\Omega)\tau}{2k\ell}+\dfrac{\ell^2}{2k\ell}(\textbf{q}^2-\textbf{q}\cdot\textbf{q}_1)\right]+\nonumber\\
&&+2ig\tau\int\!\!\!\int \dfrac{dE}{2\pi}\dfrac{d\Omega}{2\pi}
\int\dfrac{d^3\textbf{q}_1}{(2\pi)^3}
\Omega\, n_E(\textbf{q}_1,\Omega)
\partial_\epsilon n_\epsilon(\textbf{q}-\textbf{q}_1,\omega-\Omega),
\end{eqnarray}
which is Eq. (\ref{BS_eq_cd}) of the main text. Note that in deriving Eq. (\ref{BS_eq_cd2}), we kept only first-order terms in $\omega\tau$, $\Omega\tau$,  $\textbf{q}^2\ell^2$ and $\textbf{q}_1^2\ell^2$ (diffusion approximation), and in $\tau g n/(k\ell)\sim \tau g n/(\epsilon\tau)$. Terms of higher-order have been neglected. 

We now examine $n_\epsilon(\textbf{r},\omega)|_\text{e}$. The Fourier transform of Eq. (\ref{BS_eq_e}) with respect to $\textbf{r}$ reads
\begin{eqnarray}
n_\epsilon(\textbf{r},\omega)|_\text{e}&=&
2\gamma^2 g \int\!\!\!\int\!\!\!\int\dfrac{d^3\textbf{q}_1}{(2\pi)^3}\dfrac{d^3\textbf{k}}{(2\pi)^3}\dfrac{d^3\textbf{k}^\prime}{(2\pi)^3}
\int\!\!\!\int\dfrac{d\epsilon_3}{2\pi}\dfrac{d\epsilon_4}{2\pi}
\overline{G}^*_{\epsilon_2}(\textbf{k}-\textbf{q})\overline{G}_{\epsilon_1-\epsilon_3+\epsilon_4}(\textbf{k}-\textbf{q}_1)\overline{G}_{\epsilon_1-\epsilon_3+\epsilon_4}(\textbf{k}^\prime-\textbf{q}_1)\overline{G}_{\epsilon_1}(\textbf{k})\overline{G}_{\epsilon_1}(\textbf{k}^\prime)\times\nonumber\\
&&\times 
n(\textbf{q}_1,\epsilon_3,\epsilon_4)
n(\textbf{q}-\textbf{q}_1,\epsilon_1-\epsilon_3+\epsilon_4,\epsilon_2).
\end{eqnarray}
As compared to Eq. (\ref{B1}), an additional integral over $\textbf{k}^\prime$ comes into play. At the leading order in $1/(k\ell)\sim1/(\epsilon\tau)$, this integral reads
\begin{equation}
\gamma\int\dfrac{d^3\textbf{k}^\prime}{(2\pi)^3}=\dfrac{i}{2k\ell}.
\end{equation}
The remaining integral over $\textbf{k}$ is the same as the one in Eq. (\ref{B1}). Keeping only first-order terms in $\tau g n/(k\ell)$, $\omega\tau$, $\Omega\tau$, $\textbf{q}^2\ell^2$ and $\textbf{q}_1^2\ell^2$, we have
\begin{eqnarray}
&&\gamma^2\int\!\!\!\int\dfrac{d^3\textbf{k}}{(2\pi)^3}\dfrac{d^3\textbf{k}^\prime}{(2\pi)^3}
\overline{G}^*_{\epsilon_2}(\textbf{k}-\textbf{q})\overline{G}_{\epsilon_1-\epsilon_3+\epsilon_4}(\textbf{k}-\textbf{q}_1)\overline{G}_{\epsilon_1-\epsilon_3+\epsilon_4}(\textbf{k}^\prime-\textbf{q}_1)\overline{G}_{\epsilon_1}(\textbf{k})\overline{G}_{\epsilon_1}(\textbf{k}^\prime)=\nonumber\\
&&\dfrac{i\tau}{2k\ell}\left[-i+2(\omega-\Omega)\tau+\ell^2i(\textbf{q}^2-\textbf{q}\cdot\textbf{q}_1+\dfrac{\textbf{q}_1}{3})\right].
\end{eqnarray}
Using the same substitution as above to evaluate $n_\epsilon(\textbf{r},\omega)|_\text{f}$, we obtain
\begin{eqnarray}
\label{neplusnf}
&&n_\epsilon(\textbf{r},\omega)|_\text{e}+n_\epsilon(\textbf{r},\omega)|_\text{f}=\nonumber\\
&&2g\tau\int\dfrac{d^3\textbf{q}_1}{(2\pi)^3}\int\!\!\!\int\dfrac{dE}{2\pi}\dfrac{d\Omega}{2\pi}
n_E(\textbf{q}_1,\Omega)n_{\epsilon-\Omega/2}(\textbf{q}-\textbf{q}_1,\omega-\Omega)
\dfrac{i}{2k\ell}\left[-i+2(\omega-\Omega)\tau+\ell^2i(\textbf{q}^2-\textbf{q}\cdot\textbf{q}_1+\dfrac{\textbf{q}_1}{3})\right]+\nonumber\\
&&2g\tau\int\dfrac{d^3\textbf{q}_1}{(2\pi)^3}\int\!\!\!\int\dfrac{dE}{2\pi}\dfrac{d\Omega}{2\pi}
n_E(\textbf{q}_1,\Omega)n_{\epsilon+\Omega/2}(\textbf{q}-\textbf{q}_1,\omega-\Omega)
\dfrac{-i}{2k\ell}\left[i-2(\omega-\Omega)\tau-\ell^2i(\textbf{q}^2-\textbf{q}\cdot\textbf{q}_1+\dfrac{\textbf{q}_1}{3})\right].
\end{eqnarray}
We now expand the densities within the integrals, using Eq. (\ref{nepsilon_expansion}). However, this time we can restrict ourselves to the zero-order term in the expansion (\ref{nepsilon_expansion}), since keeping the term proportional to $\Omega$ would lead to terms of second-order in $\omega\tau$, $\Omega\tau$ and $\textbf{q}^2\ell^2$, which are negligible in the diffusion approximation. We thus write $n_{\epsilon\pm\Omega/2}\simeq n_\epsilon$, and obtain
\begin{eqnarray}
\label{BS_eq_ef2}
n_\epsilon(\textbf{r},\omega)|_\text{e}+
n_\epsilon(\textbf{r},\omega)|_\text{f}&=&
4g\tau\int\!\!\!\int \dfrac{dE}{2\pi}\dfrac{d\Omega}{2\pi}
\int\dfrac{d^3\textbf{q}_1}{(2\pi)^3}
n_E(\textbf{q}_1,\Omega)
n_\epsilon(\textbf{q}-\textbf{q}_1,\omega-\Omega)\times\nonumber\\
&&\times\left[\dfrac{1+2i(\omega-\Omega)\tau}{2k\ell}+\dfrac{\ell^2}{2k\ell}(-\textbf{q}^2+\textbf{q}\cdot\textbf{q}_1-\dfrac{\textbf{q}_1^2}{3})\right],
\end{eqnarray}
which is Eq. (\ref{BS_eq_ef}) of the main text.

\end{document}